\documentclass[fleqn,usenatbib]{mnras}

\usepackage[T1]{fontenc}
\usepackage{ae,aecompl}
\usepackage{color}

\usepackage{graphicx}	
\usepackage{newtxtext,newtxmath}
\usepackage{amsmath}	
\usepackage{amssymb}	

\newcommand{\Rh}{R_{\rm h}}
\newcommand{\rJ}{r_{\rm J}}
\newcommand{\vrot}{v_{\rm rot}}
\newcommand{\Nbody}{$N$-body }
\newcommand{\phiomega}{\phi_{\omega}}
\newcommand{\thetaomega}{\theta_{\omega}}
\newcommand{\thetaomegai}{\theta_{\rm \omega,i}}
\newcommand{\omegavec}{\boldsymbol{\omega}}
\newcommand{\stsr}{\sigma_{\rm T}/\sigma_{\rm R}}
\newcommand{\trhi}{t_{\rm rh,i}}
\newcommand{\rhratio}{\rm r_{h,1G}/\rm r_{h,2G}}
\setcitestyle{aysep={},citesep={,}} 

\title[Kinematics of multiple stellar populations]{Kinematical evolution of multiple stellar populations in star clusters}

\author[M. Tiongco, E. Vesperini and A.~L. Varri]{
Maria A. Tiongco,$^{1}$\thanks{E-mail: mtiongco@indiana.edu}
Enrico Vesperini,$^{1}$
and Anna Lisa Varri$^{2, 3}$
\\
$^{1}$Department of Astronomy, Indiana University, Bloomington, IN 47405, USA\\
$^{2}$Institute for Astronomy, University of Edinburgh, Royal Observatory, Blackford Hill, Edinburgh EH9 3HJ, UK\\
$^{3}$Department of Astronomy, Graduate School of Science, The University of Tokyo,  7-3-1 Hongo, Bunkyo-ku, Tokyo, 113-0033, Japan}

\date{Accepted XXX. Received YYY; in original form ZZZ}

\pubyear{2015}

\begin{document}
\label{firstpage}
\pagerange{\pageref{firstpage}--\pageref{lastpage}}
\maketitle

\begin{abstract}

We present the results of a suite of \Nbody simulations aimed at understanding the fundamental aspects of the long-term evolution of the internal kinematics of multiple stellar populations in globular clusters. Our models enable us to study the cooperative effects of internal, relaxation-driven processes and external, tidally-induced perturbations on the structural and kinematic properties of multiple-population globular clusters. To analyse the dynamical behaviour of the multiple stellar populations in a variety of spin-orbit coupling conditions, we have considered three reference cases in which the tidally perturbed star cluster rotates along an axis oriented in different directions with respect to the orbital angular momentum vector. We focus specifically on the characterisation of the evolution of the degree of differential rotation and anisotropy in the velocity space, and we quantify the process of spatial and kinematic mixing of the two populations. In light of recent and forthcoming explorations of the internal kinematics of this class of stellar systems by means of line-of sight and astrometric measurements, we also investigate the implications of projection effects and spatial distribution of the stars adopted as tracers. The kinematic and structural richness emerging from our models further emphasises the need and the importance of observational studies aimed at building a complete kinematical picture of the multiple population phenomenon.

\end{abstract}

\begin{keywords}
methods:numerical -- galaxies: star clusters: general -- Galaxy: globular clusters: general
\end{keywords}



\section{Introduction}
A new picture of globular clusters is emerging from numerous observational studies which are revealing that these stellar systems are characterised by a rich spectrum of chemical and dynamical properties.

Spectroscopic and photometric studies have shown the presence of multiple stellar populations differing in the abundances of a number of light elements \citep[see, e.g.,][]{carretta2009a,carretta2009b,piotto2015,milone2017}. At the same time, large ground-based radial velocity surveys \citep[see, e.g.,][]{bellazzini2012, fabricius2014, lardo2015, kamann2018, ferraro2018}  and astrometric studies \citep[see, e.g.,][]{watkins2015, bellini2017, bianchini2018, vasiliev2018, libralato2018,libralato2019, baumgardt2019,sollima2019}  have shown evidence of internal kinematical properties more complex than those included in the traditional view of globular clusters: many clusters are characterized by internal rotation and velocity anisotropy,  dynamical ingredients neglected in most studies aimed at modeling clusters' properties and their evolution.

A few investigations have further sharpened the dynamical picture of globular clusters by studying the kinematical properties of the multiple stellar populations within them and provided additional key constraints to understand their formation and dynamical evolution.
\citet{richer2013}, \citet{bellini2015}, and \citet{bellini2018}, using HST proper-motion measurements for, respectively, 47 Tucanae and NGC2808, and \citet{milone2018}, using Gaia proper-motion measurements of 47 Tuc, have shown that second-generation (2G) stars (defined as those with enhancements in the abundances of elements such as Na, N, and He) are characterized by a more radially anisotropic velocity dispersion than first-generation (1G) stars.\footnote{Although we use the terms first- and second-generation to refer to the different populations, we point out that according to some formation models \citep[see, e.g.,][]{bastian2013,gieles2018} all the populations form at the same time, and in those cases the two populations are more properly referred to as first- and second-population.} \citet{cordero2017} studied the rotational properties of the multiple populations of M13 and found that the extreme 2G population rotates more rapidly than the intermediate 2G.   In M5, \citet{lee2017} found that 2G stars  are characterized by a significant rotation while 1G stars show no rotation. An opposite trend was found instead  in NGC 6752: for this cluster the analysis presented in \citet{lee2018} revealed that 1G stars rotate more rapidly than 1G stars.
Finally, \citet{dalessandro2018} found a puzzling difference in the velocity dispersion of 1G and 2G stars in the outer regions of NGC6362.

Although the observational investigation of the kinematic dimension of the multiple population phenomenon is still in its very early stages, the kinematical differences emerging from the first observational studies mentioned above clearly show the dynamical richness and complexity of multiple-population clusters and can be exploited to provide key constraints for models of their formation and dynamical evolution.

On the theoretical side, the numerous questions concerning the different aspects of the origin and dynamics of multiple stellar populations are still open and at the center of much attention. 
A number of numerical investigations have explored different aspects of the dynamics of multiple-population globular clusters, such as the evolution of the structural properties of different populations \citep[see, e.g.,][]{vesperini2013,vesperini2018, mb2013, mb2016, miholics2015, fare2018}, the dynamics of binary stars \citep{vesperini2011, hong2015,hong2016,hong2019}, and the effects of dynamical evolution on the stellar mass function \citep{vesperini2018}. A few studies have started to specifically focus on the kinematical properties of multiple populations and attempted to distinguish kinematical features imprinted by the formation process from those which are instead shaped by the cluster's long-term dynamical processes \citep[see, e.g.,][]{bekki2010, bekki2011, mb2013, mb2016, hb2015, bellini2015,  dalessandro2018, hong2019}.

The goal of this paper is to explore the fundamental dynamical aspects of the evolution of the kinematical properties of multiple-population clusters and the connection between the evolution of the kinematical and structural properties. Specific attention will be devoted to the interplay between the effects of internal dynamics and those of the external tidal field of the host galaxy, and related implications on
the internal kinematics of 1G and 2G stars.
Although our \Nbody models are not aimed at a detailed comparison with observational data for specific globular clusters, we also explore how the intrinsic properties of our numerical simulations translates into corresponding projected observables, with a careful assessment of possible degeneracies and limitations in accessing the phase space of nearby star clusters.

The outline of the paper is the following: in Section \ref{methods}, we describe the method and initial conditions of our \Nbody simulations, in Section \ref{results}, we present the results of our study, Section \ref{discussion} includes a discussion of our results, and in Section \ref{conclusions} we state our conclusions.

\section{Method and Initial Conditions}
\label{methods}

We have created a suite of direct \Nbody simulations using {\sevensize NBODY6} with GPU acceleration \citep{aarseth2003,nitadori2012}, and ran them on the Big Red II cluster at Indiana University.

The star clusters in our simulations move on circular orbits about the centre of their host galaxy. The potential of the host galaxy is modeled as that of a point-mass, and  the equations of motions are solved in a co-rotating reference frame centred on the cluster \citep{heggie2003}, rotating with angular speed, $\Omega$, equal to that of the angular speed of the cluster's orbital motion.  The orbital plane is set up to be parallel to $x$-$y$ plane, and the cluster's orbital angular velocity vector is parallel to the $z$-axis.  All 
initial conditions begin with $N=$ 32~768 equal-mass stars, and stars are removed from the 
\Nbody system once they move beyond a distance equal to two times the cluster's Jacobi radius, $\rJ$.

Our systems start with an equal amount of second-generation (2G) and first-generation (1G) stars. Since our investigation is focused only on the long-term evolution of multiple-population star clusters,  we do not study here the early stages of a cluster's evolution which might have driven the number ratio of 2G to 1G stars toward values close to those observed in Galactic clusters \citep[see, e.g.,][]{decressin2007,dercole2008}.

Following the prediction of a number of multiple-population formation 
scenarios we start with a 2G  initially more compact than the 1G;
our systems are characterized by internal rotation, and as predicted by the hydrodynamical simulations of \citet{bekki2010,bekki2011}, we assume that the 2G is initially rotating more rapidly than the 1G population.

To set up our initial conditions, we produce the relevant rotating components from \Nbody realizations of the models by \citet{varri2012}.  These distribution function-based equilibria
are characterized by differential rotation, 
such that the rotation curve rises approximately linearly with radius in the cluster's inner regions, reaches a peak in the intermediate parts, and then decreases and vanishes at the edge of the system. These 
configurations are oblate and flattened along a direction parallel to the rotation axis, and their rotational and structural properties are determined by four dimensionless parameters, $W_0, \hat{\omega}, \bar{b}$, and $c$, which define the central concentration, the rotation strength, and the shape of the rotation curve in the intermediate and outer regions of the models, respectively \citep[see][for further details on the properties of these models]{varri2012}.
The specific parameters we have adopted for our initial conditions are the following: for the 1G system, $W_0=3$, $\hat{\omega}=0.1$, $\bar{b}=0.1$, and $c=1$; for the 2G system $W_0=7$, $\hat{\omega}=0.3$, $\bar{b}=0.1$, and $c=1$. The two models have been scaled so that the initial ratio of the 1G to the 2G half-mass radius is equal to 5.  
The choice of the values attributed to the dimensionless parameters has been informed by a number of previous phenomenological studies in which such a family of models has been used to interpret the structural and kinematic properties of Galactic globular clusters \citep[e.g., see][]{bianchini2013,kacharov2014, bellini2017}.  Similarly, the value of the ratio of the half-mass radii assumed here has been based on the current empirical knowledge of the spatial distribution of multiple populations in Galactic globular clusters, as based on existing photometric studies \citep[e.g.,][]{lardo2011}. In all cases, the values adopted here should be considered only as representative, as a global modeling exercise (based on multi-component distribution function-based equilibria) of multiple-population star clusters has not been conducted yet in the current literature.

Similar to \citet{tiongco2018}, we present the analysis of our results using an inertial reference frame.  In our analysis, we will refer to the global angular velocity vector, $\omegavec$, calculated from an entire system of particles, and we define this vector to be the rotation axis of the system.  We define the rotation axis's direction in space using the angles $\phiomega$ and $\thetaomega$: $\phiomega$ is the angle between the projection of $\omegavec$ in the $x$-$y$ plane (the orbital plane) and the $x$-axis, measured in a counter-clockwise direction from the $x$-axis, and $\thetaomega$ is the angle between $\omegavec$ and the $z$-axis, measured starting from the $z$-axis (the axis pointing outwards from the orbital plane).

Table~\ref{tab:table1} lists the simulations in this study and the initial orientation of the rotation axis of the cluster, $\thetaomegai$ (all models have their initial rotation axis in the $x$-$z$ plane).  The parts of the cluster that intrinsically have zero rotation (the outermost regions) are not assumed to be tidally locked, i.e., the outermost regions initially have zero rotation in the inertial reference frame.

We initialize each \Nbody system such that the ratio of the cluster's truncation radius is equal to the Jacobi radius, making its ratio of the half-mass radius to Jacobi radius ($r_{\rm h}/r_{\rm J}$, i.e., the cluster's initial filling factor) equal to 0.106.  We evolve each simulation until about 20--25\% of the cluster's initial mass is left.

\begin{table}
\caption{Nomenclature of the \Nbody simulations presented in this study. The symbol $\thetaomegai$ denotes the initial value of the angle between the direction of the global angular velocity vector and the $z$-axis, measured starting from the $z$-axis (which points outwards from the orbital plane). For further details, see Section \ref{methods}.}
\label{tab:table1}
\begin{tabular}{@{}cccc}
\hline
Model ID & 
$\thetaomegai$ (degrees) \\
\hline
Theta0 & 0  \\
Theta45 & 45  \\
Theta90 &  90  \\
\hline
\end{tabular}
\end{table}

\section{Results}
\label{results}
\subsection{Rotation}
\subsubsection{Case of the rotation axis parallel to the orbital angular velocity vector}

\begin{figure*}
	\includegraphics[height=5in]{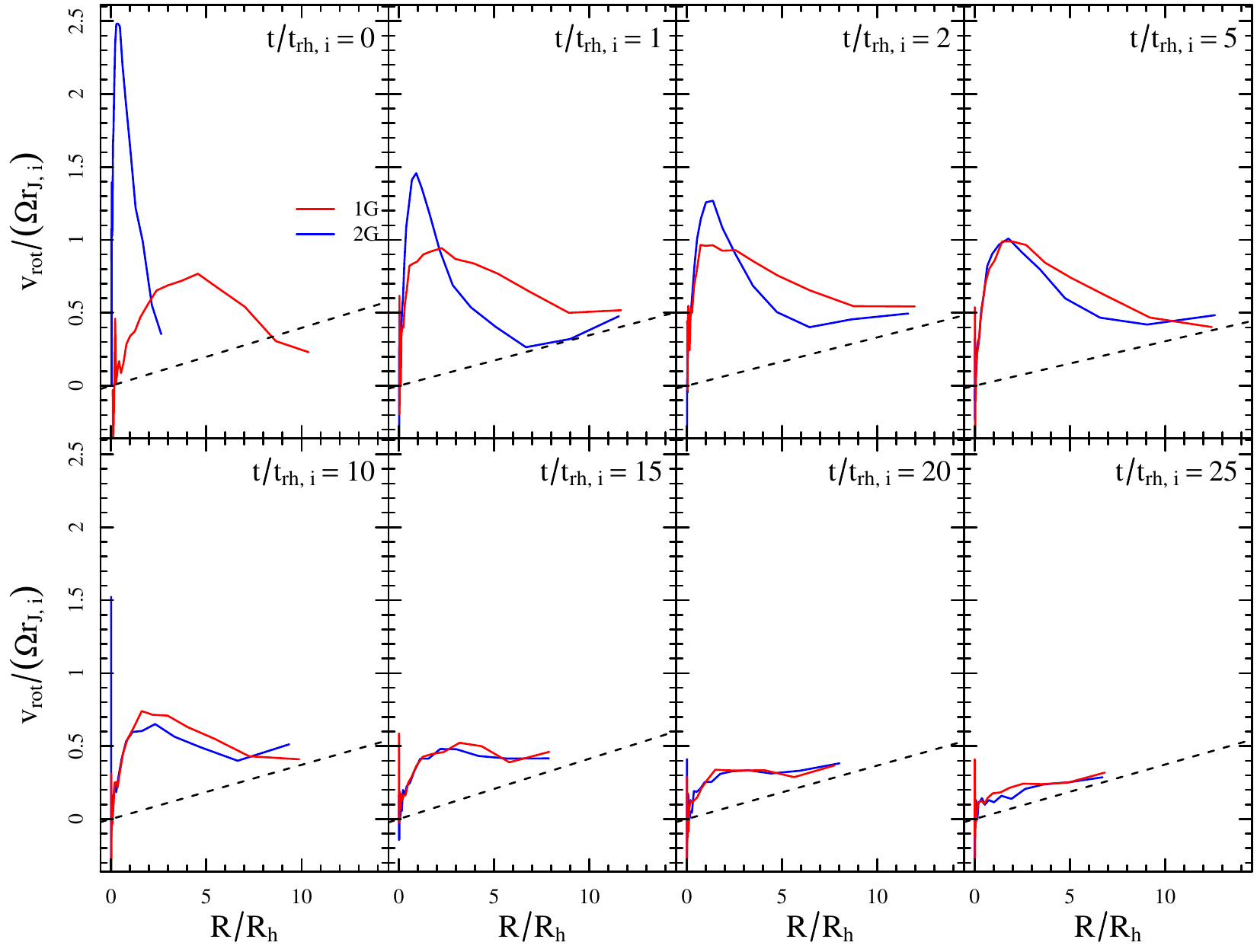}
    \caption{Time evolution of the rotation curves for each population in model Theta0, in which the rotation axis is parallel to the orbital angular velocity vector. The projected 2D radius $R$ is normalized to the projected half-mass radius, $\Rh$ of the entire system. The rotational velocity, $\vrot$, is measured about a line of sight parallel to the rotation axis (i.e., the $z$-axis), and is normalized to $\Omega r_{\rm J,i}$, the speed of co-rotation with the cluster's orbital motion at the initial Jacobi radius.  Time is normalized to the initial half-mass relaxation time of the whole cluster, $\trhi$.  The dashed line represents a solid-body rotation curve with slope 0.5$\Omega$ (see main text for details).}
    \label{fig:vrotevol0}
\end{figure*}

\begin{figure*}
	\includegraphics{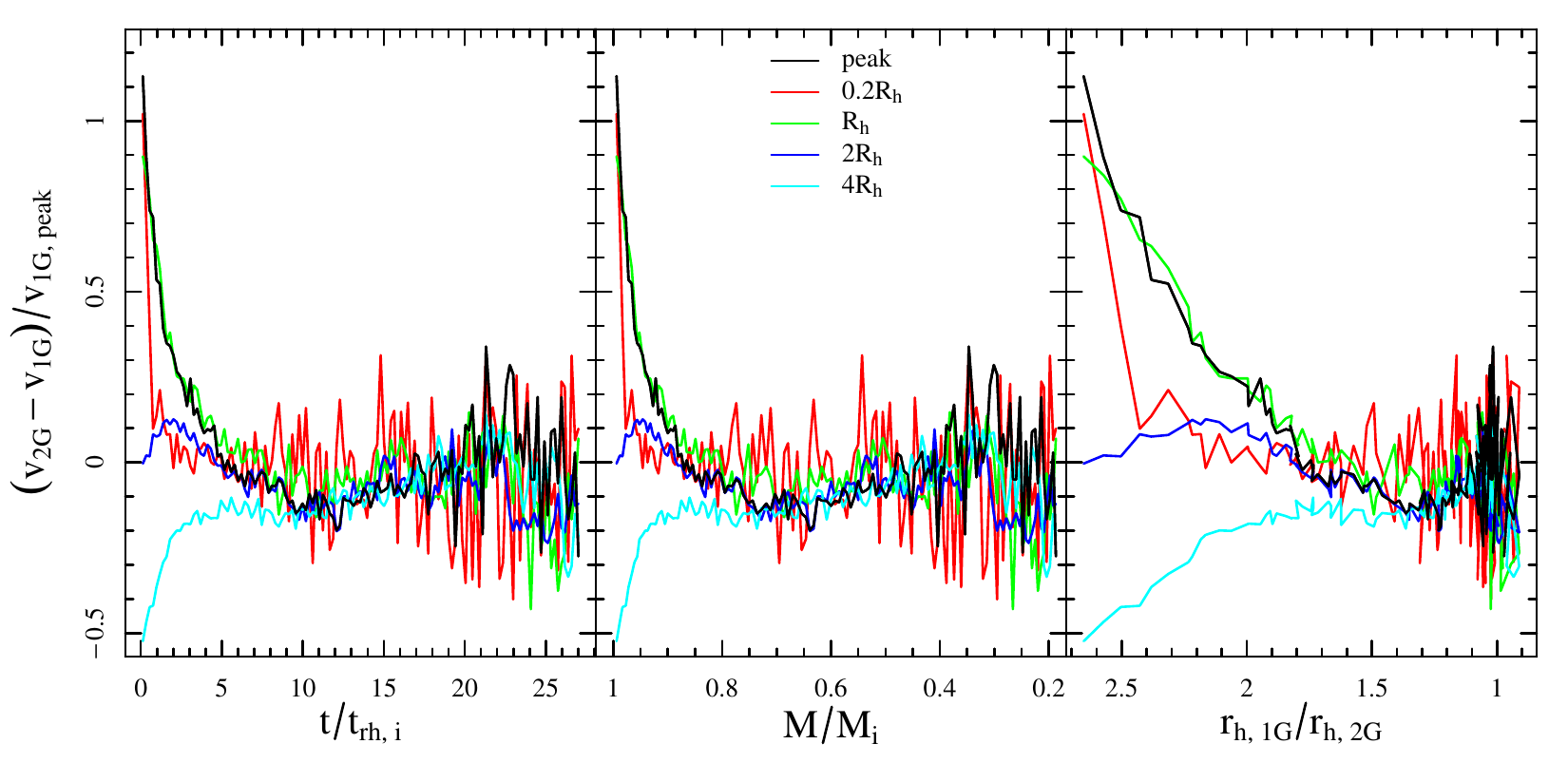}
    \caption{Evolution of the difference in rotational velocity of the two populations at different locations within the cluster for model Theta0, normalised to the peak velocity of 1G.  Left panel: evolution as a function of time, normalized to the initial half-mass relaxation time. Middle panel: evolution as a function of fraction of initial mass remaining in the cluster. Right panel: evolution as a function of the ratio of the 3D half-mass radius of each population.}
    \label{fig:rotdiff}
\end{figure*}

\begin{figure*}
	\includegraphics{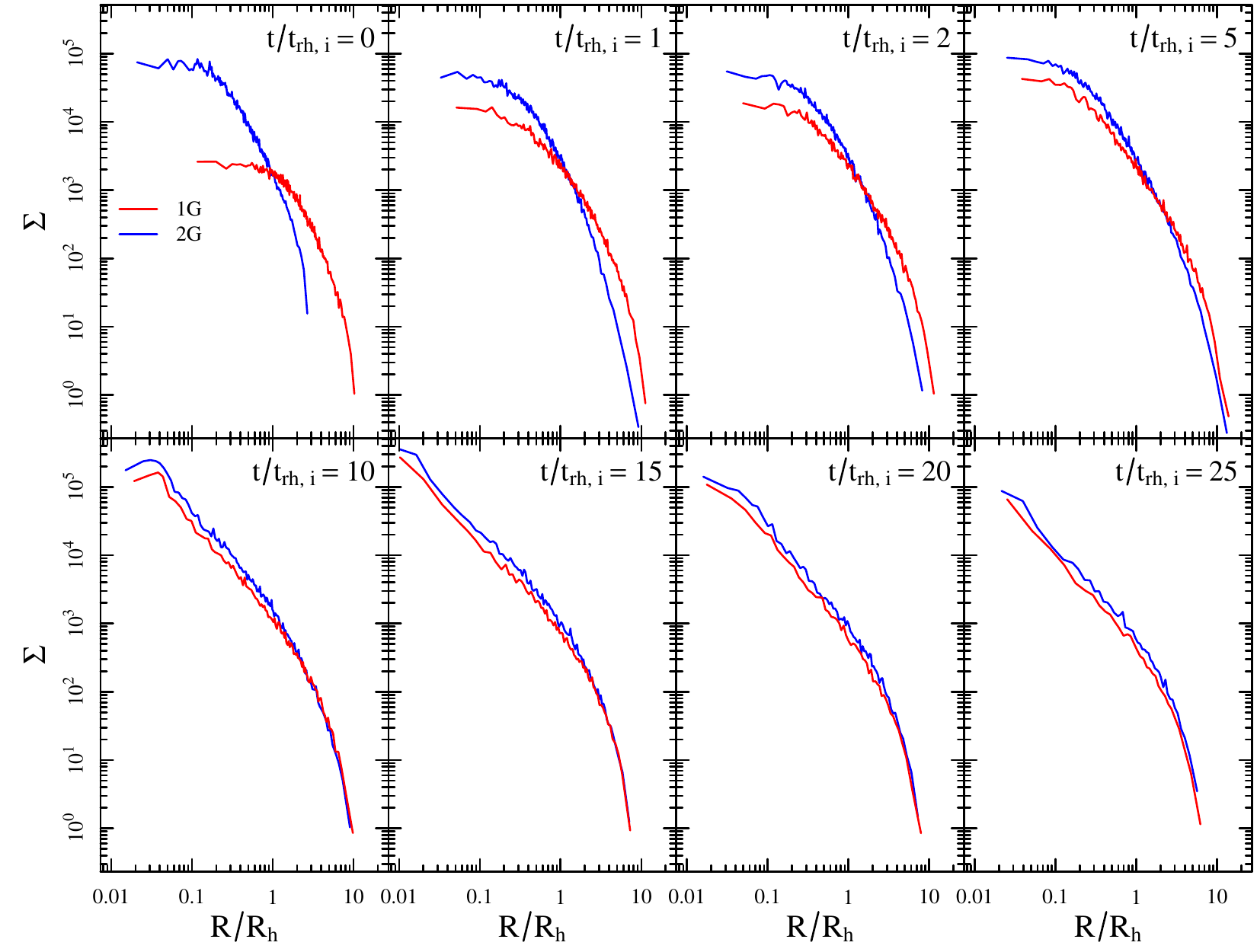}
    \caption{Time evolution of the projected number density profiles for each population in model Theta0, where the line of sight is the same as Fig.~\ref{fig:vrotevol0}.}
    \label{fig:denevol}
\end{figure*}

We begin the presentation of our results by considering first the simplest case among those under consideration (see Table 1). Therefore, we focus our attention on  the evolution of the rotational properties of model Theta0, in which we assume that the internal and orbital angular velocity vectors are initially aligned \citep[e.g., see][]{ernst2007,kim2008}.  Later, we will discuss the evolution of the radial distribution of the angular momentum
in the more general case,  where the internal and orbital angular velocity vectors are \textit{not aligned}. We will then move onto the discussion of the evolution of the degree of anisotropy of the velocity dispersion tensor in our \Nbody models.

Fig.~\ref{fig:vrotevol0} shows the evolution of the rotational velocity profiles of the two populations for the Theta0 model. We adopt a line of sight parallel to the rotation axis of the system (the $z$-axis) and the rotational velocity profile (as a function of the cylindrical radius) is calculated from the tangential components of the 
velocity vectors of the particles, as projected on the plane perpendicular to the chosen line of sight.    The profiles are constructed by combining 5 snapshots around the desired time {(except for time $t/\trhi=0$, where only 1 snapshot is used)}, taking into consideration all particles within the Jacobi radius and binning them in cylindrical shells with heights encompassing the entire cluster.

In agreement with previous studies \citep[see e.g.,][]{einsel1999,ernst2007,kim2008,tiongco2017}, we find that the overall cluster's rotation decreases due to two-body relaxation redistributing angular momentum from the inner regions to the outer regions and the loss of angular momentum carried away by escaping stars.

The effects of two-body relaxation erase the initial differences between the spatial and kinematical properties of the two populations. The profiles shown in Fig.~\ref{fig:vrotevol0} illustrate how the rotational velocity profiles of the two populations evolve and become increasingly similar.  Specifically, the 2G population's peak rotational velocity decreases and its location moves to higher fractions of $\Rh$.  Meanwhile, the changes in the rotation curve of 1G stars are more subtle and the location of its peak velocity has only a moderate shift toward the inner regions.
We point out that, although over most of the cluster's extension the 2G population is characterized by a larger rotational velocity than the 1G population, as  2G stars diffuse out of the central core and start populating the cluster's outermost regions, their outer rotational velocity may be slower than that of the 1G population. This transition in the difference between the 1G and the 2G rotational velocities is therefore a kinematic manifestation and a consequence of the initial structural differences of the two populations.

After the rotational profiles of the two populations become indistinguishable, they  share the same evolution towards smaller values of the rotational velocity, eventually converging towards a solid-body rotation, in agreement with what found in \citet{tiongco2016b,tiongco2017}. As shown in \citet[][]{tiongco2016b}, the combined effects due to the external tidal field and the preferential loss of prograde orbiting stars result in  an internal rotation only \textit{partially} synchronized with the external orbital rotation with an internal angular velocity $\omega \simeq 0.5 \Omega$ \citep[see also][]{claydon2017}.  In Fig.~\ref{fig:vrotevol0}, we also show with a dashed line the solid-body rotation corresponding to $\omega = 0.5\Omega$, to illustrate how the rotation curve is the combination of the intrinsic rotation curve (that is gradually being erased) and the growing solid-body rotation due to the effects of the external tidal field on the cluster's kinematics  (though the value of this slope can not be disentangled from the composite rotation curve until the final stages of dynamical evolution).  The final panel of Fig.~\ref{fig:vrotevol0} shows that there is still some intrinsic rotation left in the cluster by this point in the simulation, not yet converging to $\omega \simeq 0.5\Omega$.

Further details of the evolution of the rotational properties of the 1G and the 2G populations are presented in Fig.~\ref{fig:rotdiff}; this figure shows the evolution of the difference between 1G's and 2G's rotational velocity measured at different radii from the cluster centre, and at the peak rotational velocity of each population. This figure illustrates how the difference between the 1G and the 2G rotational velocity evolves as a function of three different measures of dynamical evolution within the cluster: (i) time normalized to the initial half-mass relaxation time, (ii) the fraction of the initial mass left in the cluster, and (iii) the ratio of 3D half-mass radius of 1G to the 3D half-mass radius of 2G, $\rhratio$.

The picture emerging from this figure further illustrates how initial differences in the 1G and 2G rotational properties are gradually erased during a cluster's evolution: the inner regions of the cluster are those where the local relaxation time is shorter and where memory of the initial rotational differences is lost more rapidly. The right panel of Fig.~\ref{fig:rotdiff}  shows the evolution of the  rotational difference as a function of $\rhratio$; this ratio is a global measure of the spatial mixing of the two populations and allows us to illustrate the connection between  the spatial and kinematical mixing of the two populations. It is interesting to point out that the mixing in the rotational profiles can be reached before complete spatial mixing. We illustrate this point further in Fig.~\ref{fig:denevol}, by showing the evolution of the projected number density profiles ($\Sigma(R)$) featuring the same times as Fig.~\ref{fig:vrotevol0} (however, it should be noted that the residual spatial radial gradient remaining when the two populations are kinematically mixed is very weak and likely difficult to observe).

\subsubsection{Case of the rotation axis at a generic angle to the orbital angular velocity vector}

\begin{figure*}
	\includegraphics[height=4.4in]{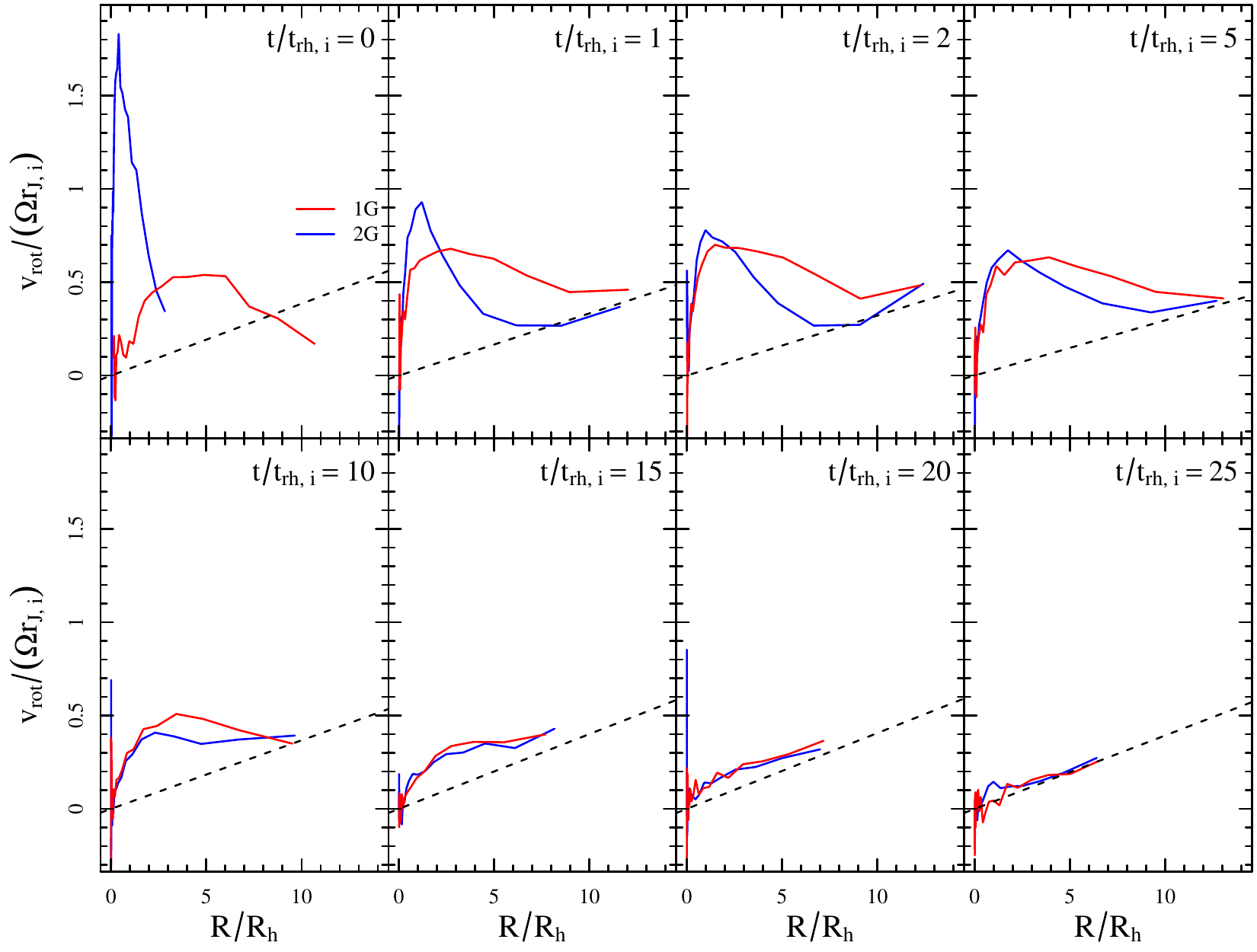}
    \includegraphics[height=4.4in]{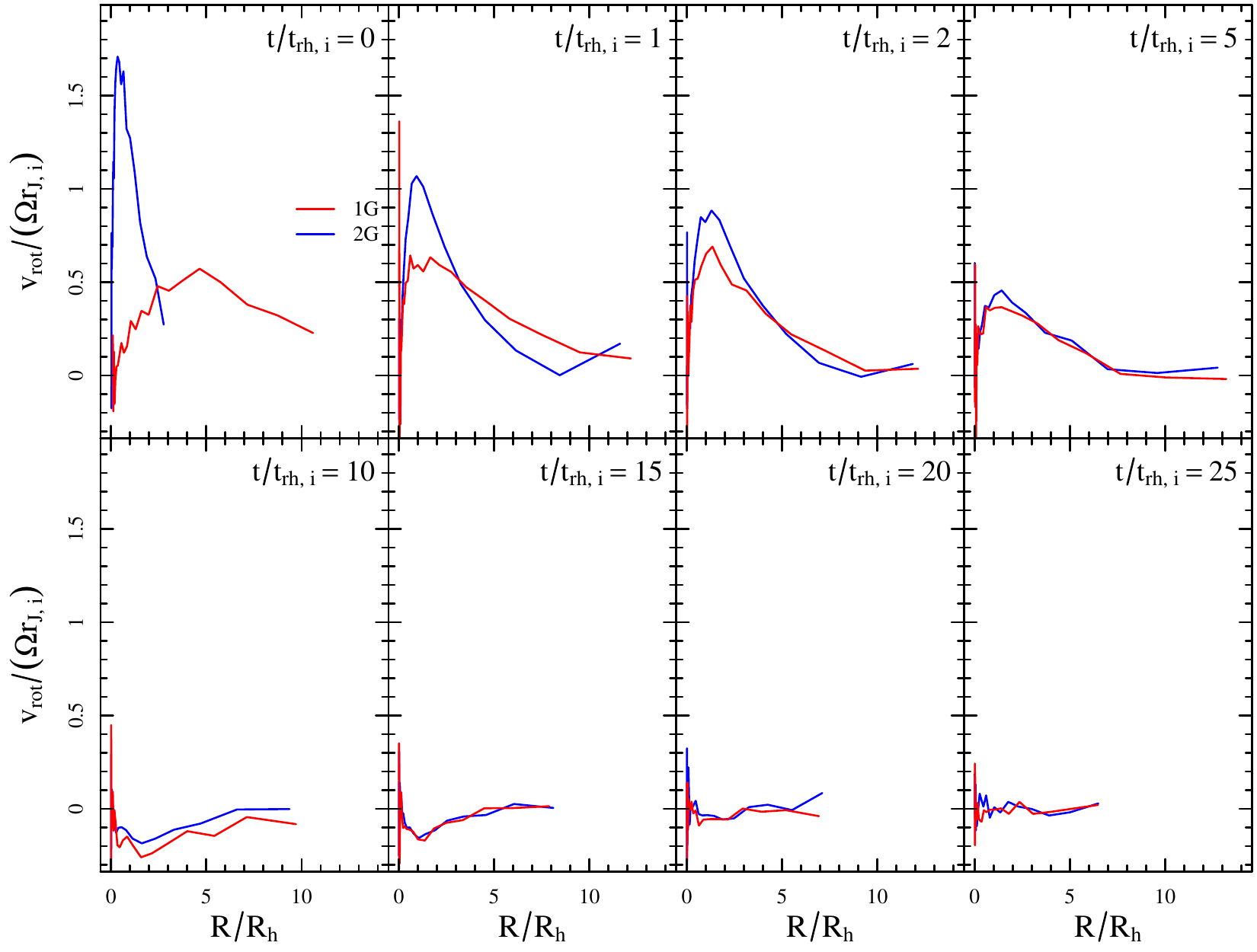}
    \caption{Evolution of the rotation curves for each population in model Theta45, in which the initial rotation axis is at a generic angle to the orbital angular velocity vector. Notation and line colours  are consistent with those adopted in Fig.~\ref{fig:vrotevol0}. The rotational velocity, $\vrot$, is measured about the $z$-axis (top two rows), and then the $x$-axis (bottom two rows).}
    \label{fig:vrotevol45}
\end{figure*}

\begin{figure*}
	\includegraphics{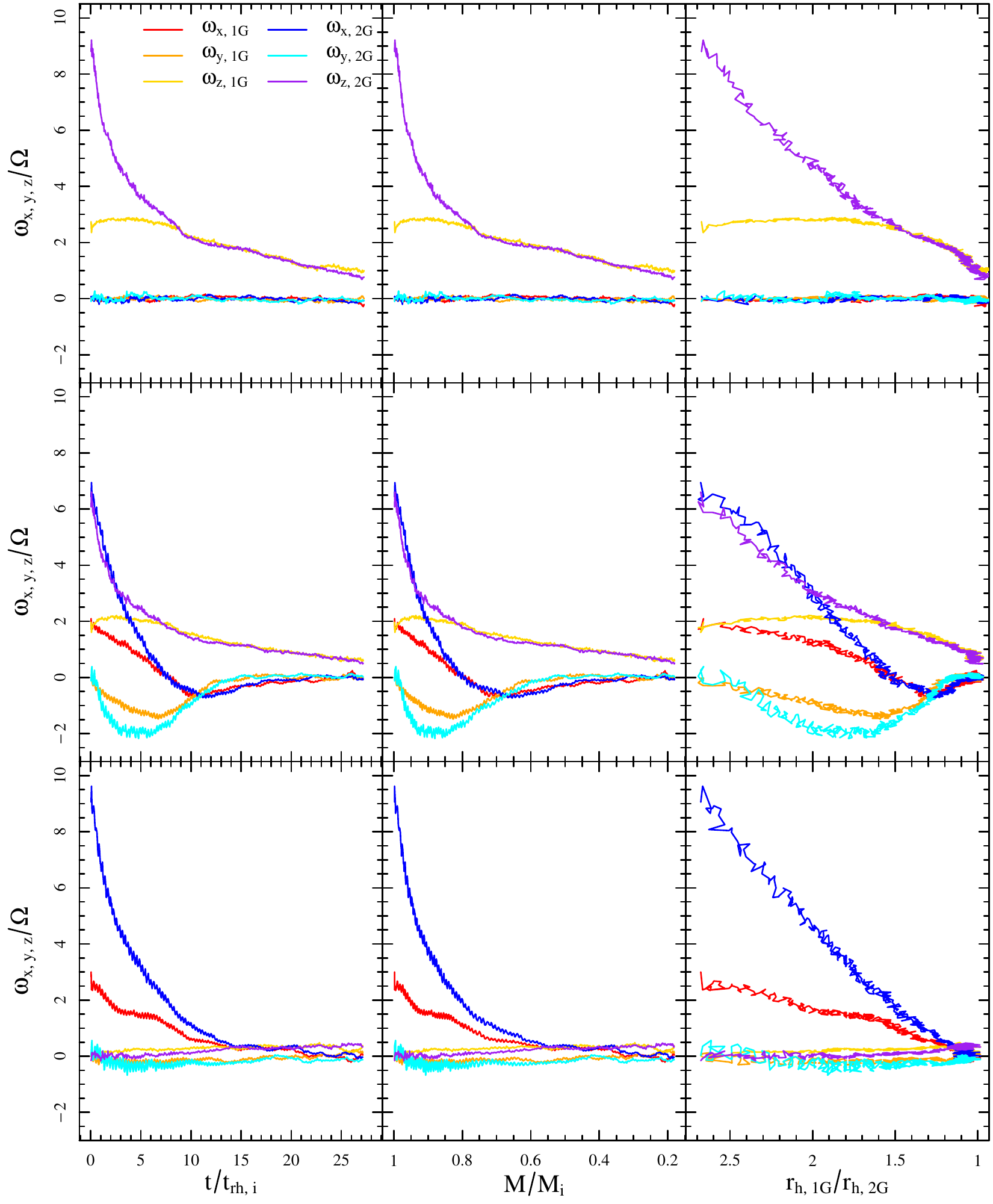}
    \caption{Evolution of the Cartesian components of the global angular velocity vector, $\omegavec$ of each population in model Theta0 (top row), Theta45 (middle), and Theta90 (bottom), normalised to the angular speed of the cluster's orbital angular velocity, $\Omega$.  The notation adopted for the quantities illustrated on the horizontal axis is described in Fig.~\ref{fig:rotdiff}.}
    \label{fig:omglob}
\end{figure*}

\begin{figure}
	\includegraphics{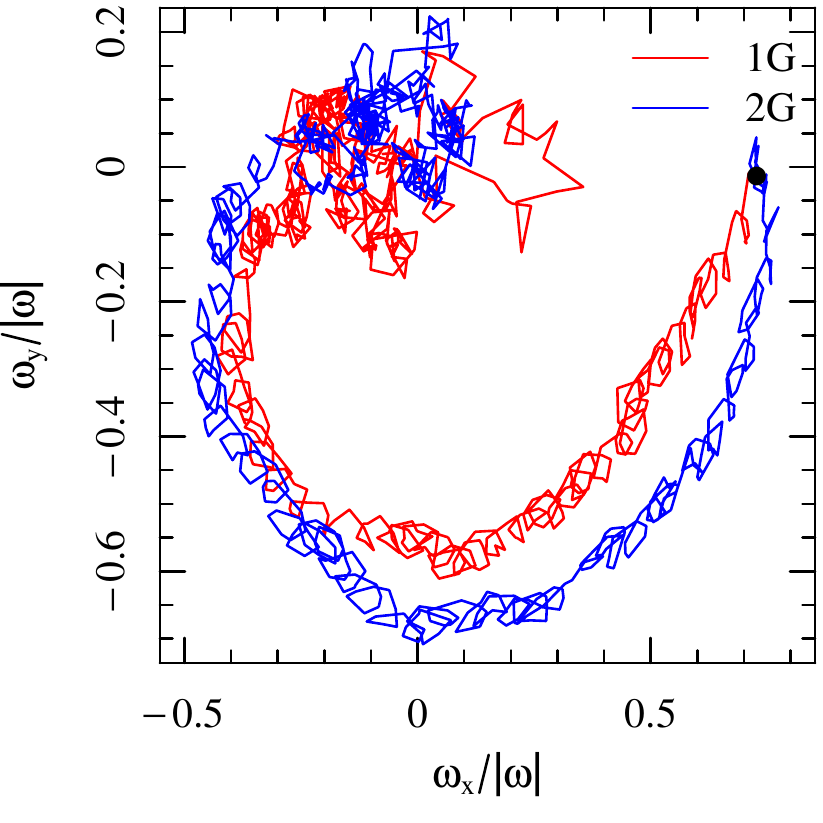}
    \caption{Evolution of the Cartesian components $\omega_{\rm x}$ and $\omega_{\rm y}$ of each population in model Theta45. The components are normalized to the magnitude of the vector $\omegavec$, calculated at that time.  The black dot near the top right of the figure indicates the starting point of the time sequences.}
    \label{fig:omxvomy}
\end{figure}

\begin{figure*}
	\includegraphics{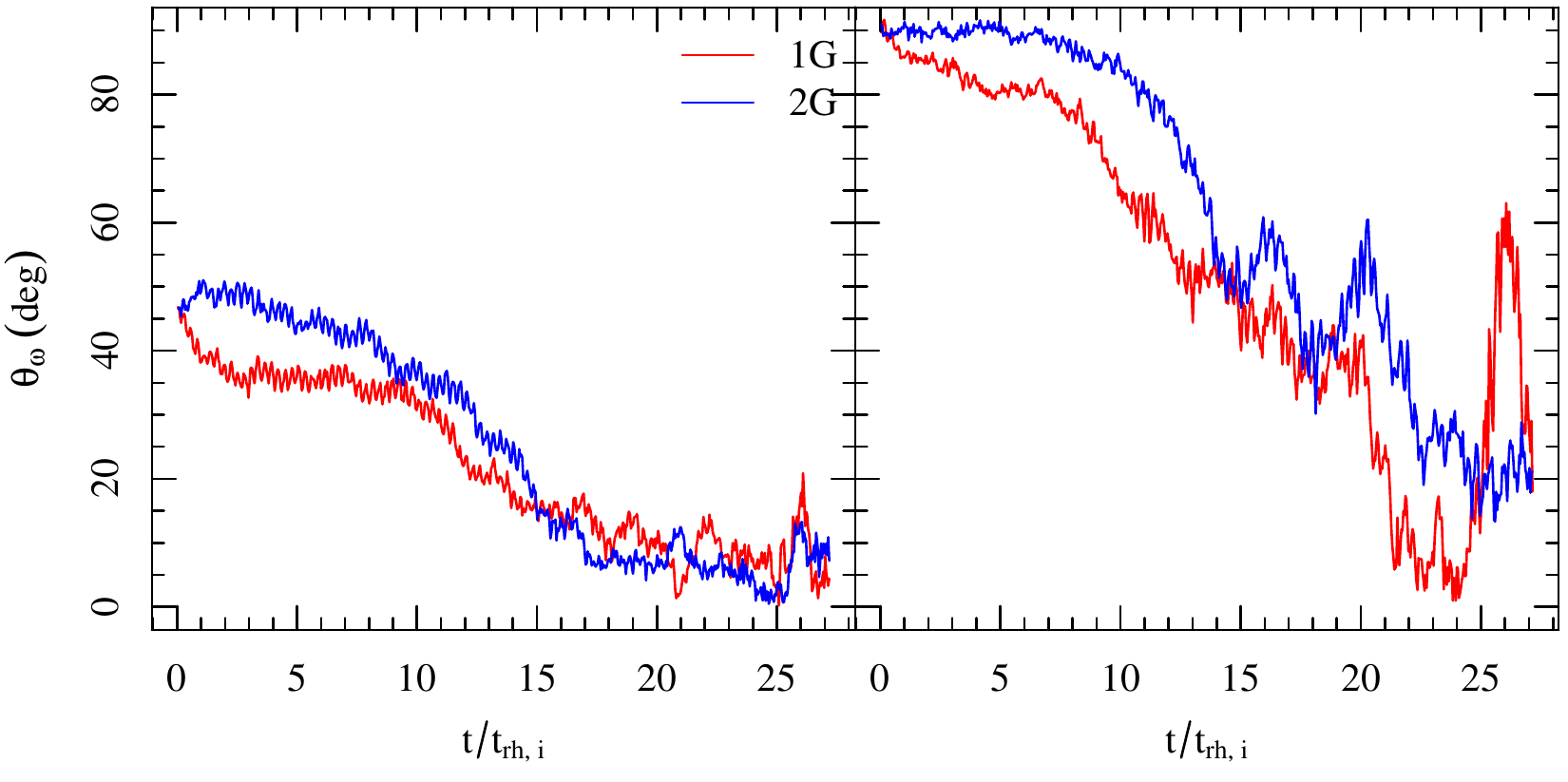}
    \caption{Evolution of $\thetaomega$, the angle in degrees between $\omegavec$ and the $z$-axis (pointing outward perpendicular to the orbital plane) for each population in model Theta45 (left) and Theta90 (right).  The evolution is plotted as a function of time normalised to the initial half-mass relaxation time, $\trhi$}
    \label{fig:thetaomega}
\end{figure*}

\begin{figure}
	\includegraphics{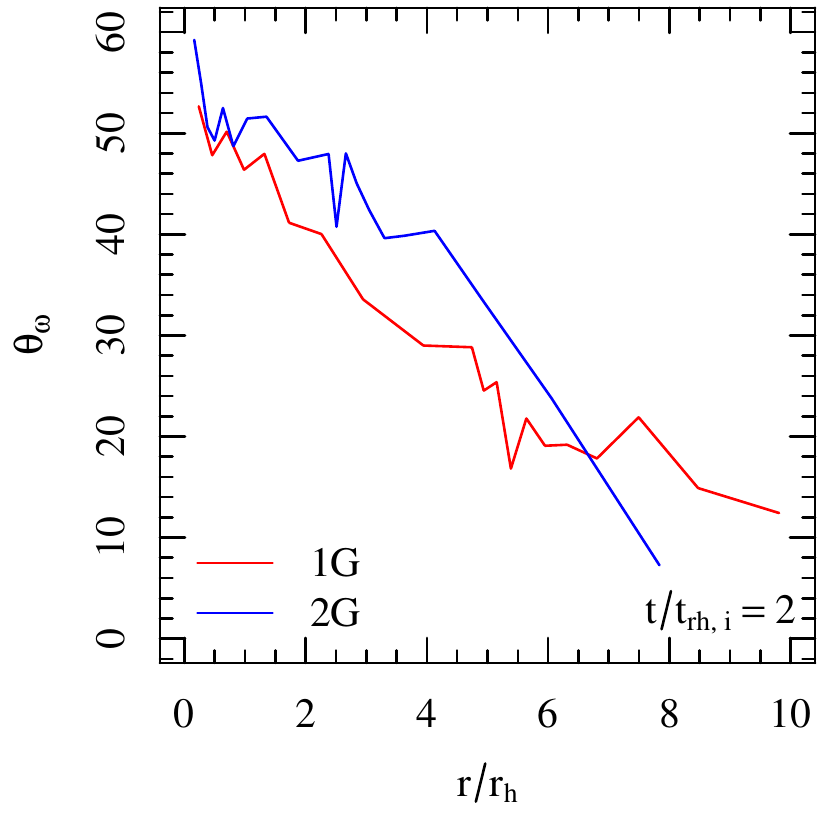}
    \caption{{The radial profile of $\thetaomega$ for each population in model Theta45 at time $t/\trhi = 2$.  Radius is normalized to the 3D half-mass radius of the entire system at that time.}}
    \label{fig:thetaomvsr}
\end{figure}

Building on the understanding developed in the simpler case described in the previous subsection, here we follow the evolution of the kinematical properties of the \Nbody models in which we have relaxed the assumption concerning the initial alignment of the internal and orbital angular velocities. We now consider the case of a stellar system in which the initial internal rotation axis points in a generic direction with respect to the one defined by the orbital angular velocity vector.
In \citet{tiongco2018}, we already showed that tidally perturbed rotating stellar systems characterized by an initial rotation axis oriented in a generic direction can develop a number of  complex kinematic features, such as precession and nutation of the cluster's rotation axis, a radial variation in the rotation axis' orientation  within the cluster, and counter-rotation between different parts of the cluster.  As a more realistic extension of our physical set-up, we now explore how the initial differences in the structural properties of the 1G and the 2G populations affect the extent and development of our previously found features for the two stellar populations within the cluster model.

We first present in Fig.~\ref{fig:vrotevol45} the evolution of the rotation curves for each population of model Theta45.  This figure demonstrates the importance of projection effects in the characterization of the rotation curve of the stellar system, as observed from different directions adopted for the line of sight.
 In the top rows, we consider a line of sight along the $z$-axis, which is the 
most favourable direction to detect the effects of (partial) tidal synchronization.  Similar to Fig.~\ref{fig:vrotevol0}, we denote with a dashed line the behaviour corresponding to $\omega = 0.5\Omega$.  Since model Theta45 begins with a smaller value of $\vrot$ along the $z$-axis than model Theta0, we can see the Theta45 model converging to the expected partial synchronization at $t/\trhi=25$, as shown in the final panel.  In the bottom rows of Fig.~\ref{fig:vrotevol45}, we consider a line of sight along the $x$-axis. Such a direction is indeed the least favourable one to characterize the rotational properties in the late evolution; unsurprisingly, we do not see the effects of (partial) tidal synchronization, therefore the rotation curve goes to zero in the outermost regions.  The precession of the rotation axis causes the value of $\vrot$ along the $x$-axis to become negative around $t/\trhi=10$.  Eventually, all signatures of rotation manifested along this line of sight are erased, as shown in the final panel.

Since precession can cause the direction of the rotation axis to vary greatly 
throughout the dynamical evolution of the tidally perturbed stellar system,  
we also show a more general characterization of the time evolution of a cluster's rotational properties, no longer restricted to a specific line of sight.  In Fig.~\ref{fig:omglob}, we show the evolution of the components of global angular velocity vector $\omegavec$ for each population in all of our models.  To obtain $\omegavec$, we calculate the moment of inertia tensor, $\mathbfit{I}$, and the angular momentum vector, $\mathbfit{L}$, and solve the equation $\mathbfit{L} = \mathbfit{I} \omegavec$.  The study of the vector $\omegavec$ allows us to focus on the cluster's actual rotational properties, including the direction of rotation at any given time. We will further discuss the manifestation of the complex kinematical features of our clusters along different lines of sight and the associated projection effects in Section \ref{discussion}.

Similar to the results of \citet{tiongco2018}, the large-scale oscillations taking place over several relaxation times in the $x$ and $y$ components indicate that the rotation axis of each population is precessing. Such features are most prominent for model Theta45. Also present are smaller oscillations (taking place over fractions of relaxation times), indicating nutation of the rotation axis.  The precession originates from the torque generated by the tidal field acting on the rotating, oblate cluster, which is initially oriented in a generic direction with respect to the symmetry introduced by the host galaxy's tidal field.  In Fig.~\ref{fig:omxvomy}, we further illustrate the precession and nutation of the rotation axis by plotting the evolution of the $x$ and $y$ components of $\omegavec$ for model Theta45.

Fig.~\ref{fig:omglob} also illustrates the gradual loss of angular momentum from each cluster model. As a result, the strength of rotation in all components decreases and the precession oscillations are damped.  In the late stages of the long-term dynamical evolution of both models Theta45 and Theta90, the system shows a small residual rotation in the $z$-component. As discussed in the previous subsection, this feature corresponds to the solid-body rotation induced by the cluster's interaction with the external tidal field (we wish to note that model Theta0 is still preserving some degree of its initial intrinsic rotation).

After all previous phases of investigation, we are now in the position to appreciate the evolution of the kinematical \textit{differences} of the two populations: Fig.~\ref{fig:omglob} also shows the time evolution of the components of the vector $\omegavec$ for the 1G and the 2G populations, separately.
A key result emerging from the \Nbody models that are oriented in a generic direction is that during the long term evolution of the system, the rotation axes of the 1G and  2G populations become misaligned.  The difference in the direction of the 1G's and 2G's rotation axes is demonstrated in Fig.~\ref{fig:thetaomega}, where we show for models Theta45 and Theta90 the evolution of $\thetaomega$ of each population.

The initial divergence in the directions of the rotation axes of each population (as seen in Fig.~\ref{fig:thetaomega}) is due to the 1G population being initially more extended and tidally filling; this implies that the interaction with the external tidal field and its effect in driving the cluster's rotation towards a partially synchronized rotation around the direction perpendicular to the cluster's orbital plane (the $z$ direction) acts more efficiently on the 1G  population. As the two populations mix spatially and kinematically (at $t/\trhi \approx$ 15--20), the rotation axes of each population become aligned again and eventually converge to a condition of partially synchronized rotation about the $z$-axis (and thus closer to $\thetaomega = 0$). 

As we have shown in \citet{tiongco2018}, the non-trivial interplay between several internal dynamical processes and the effects of the external tidal field can lead to a number of complex kinematical features, including a radial variation in the orientation of the rotation axis.  We show in Fig.~\ref{fig:thetaomvsr} that both populations in the cluster evolve to develop a radial variation of the direction of the rotation axis (i.e., a variation of $\thetaomega$ with radius); interestingly, multiple-population clusters are characterized by an additional complexity as the two populations do not follow the same radial profile of $\thetaomega$. More specifically, the 1G's rotation axis tends to point in a direction closer to that of the cluster's orbital angular velocity vector (i.e. closer to 0 degrees) across a broader range of radii. As already discussed for Fig.~\ref{fig:thetaomega}, this is a consequence of the fact that the more extended spatial distribution of 1G stars causes this population to be more significantly affected by the external tidal field.

\subsection{Velocity dispersion anisotropy}

\begin{figure*}
	\includegraphics[height=4.3in]{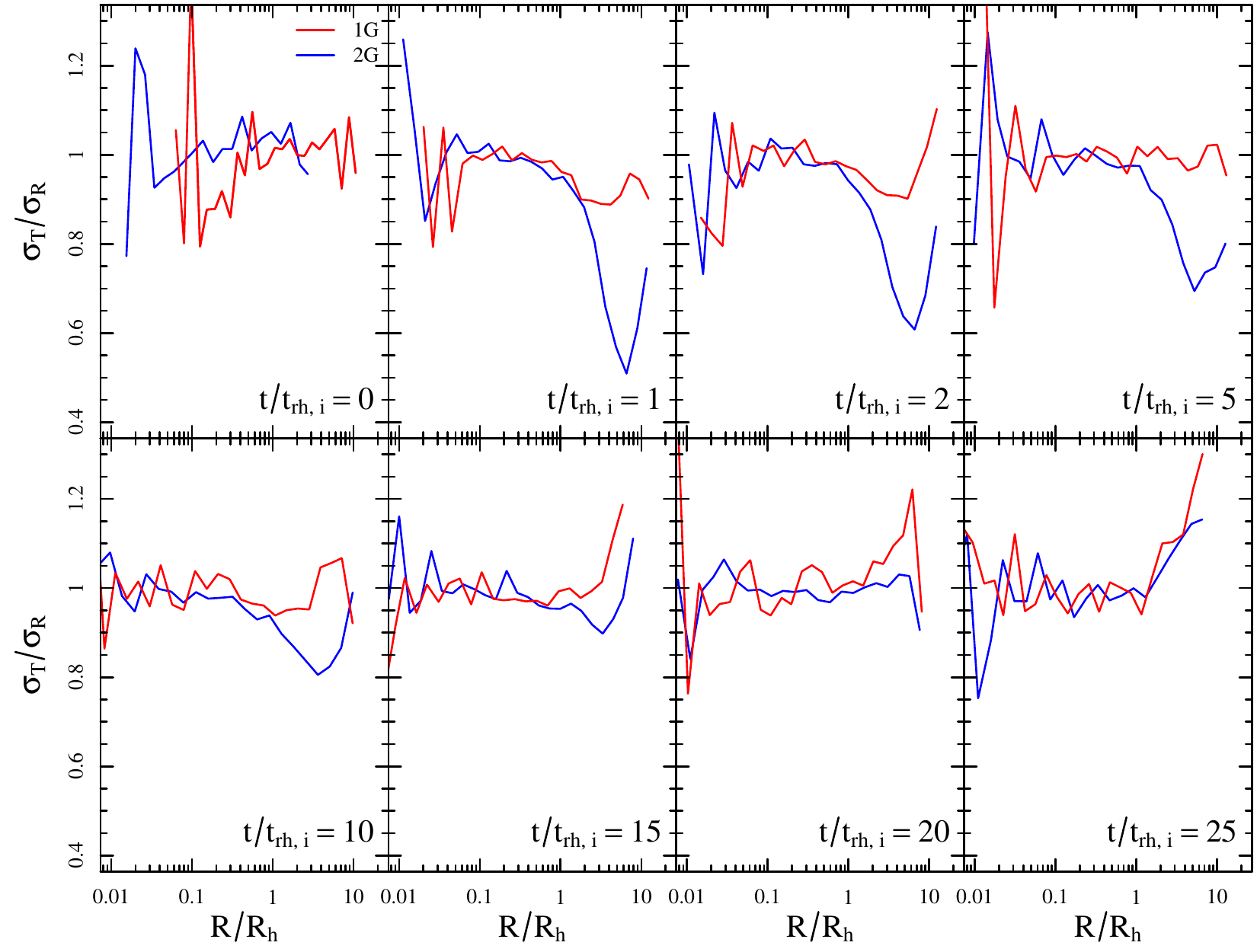}
    \includegraphics[height=4.3in]{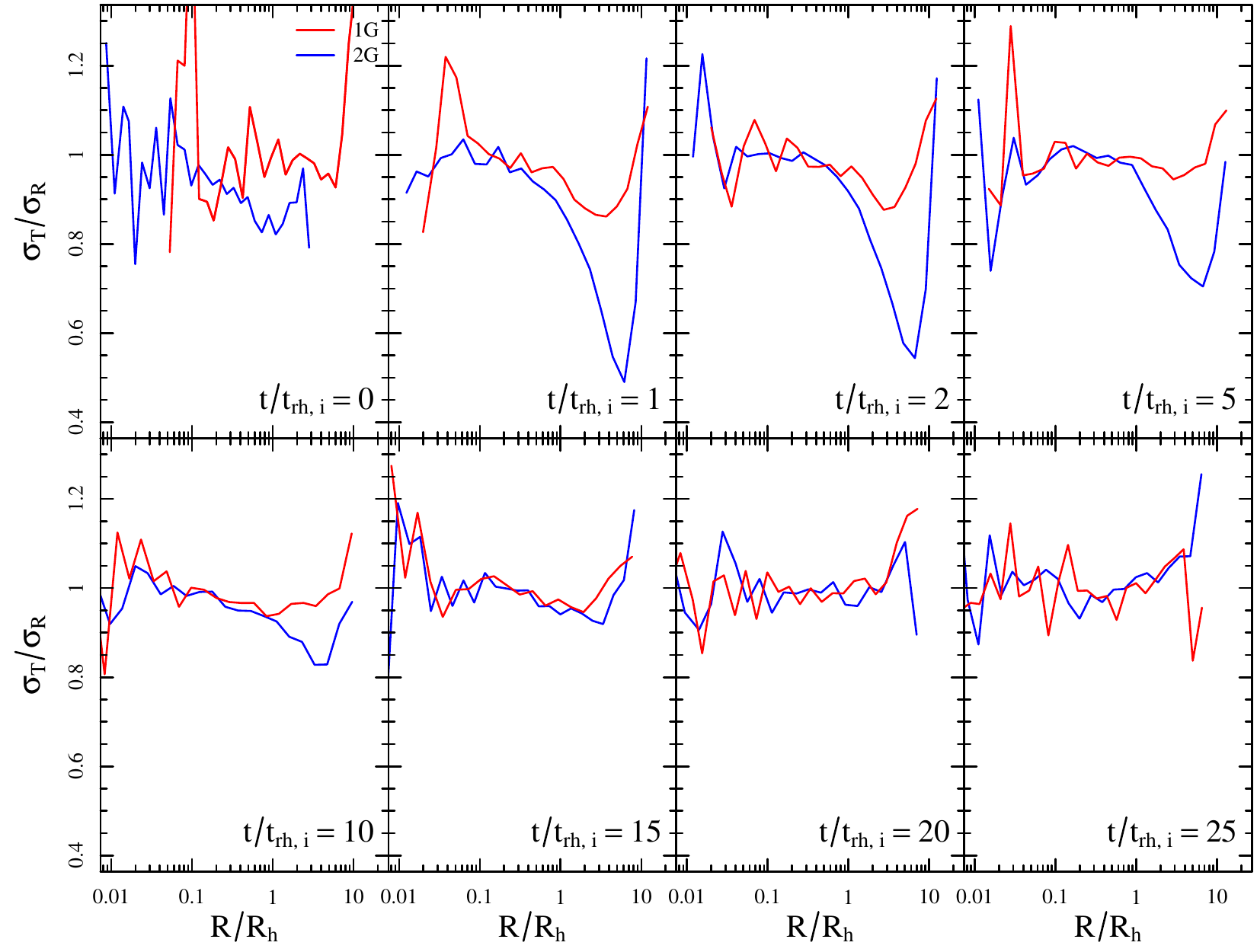}
    \caption{Time evolution of the radial profiles of the velocity anisotropy of each population in models Theta0 (top) and Theta45 (bottom).  The anisotropy parameter adopted is the ratio of the tangential velocity dispersion to the radial velocity dispersion, $\stsr$, both measured from stellar velocities projected onto the plane perpendicular to the line of sight.  The line of sight adopted is the $x$-axis in the top panel and the $z$-axis in the bottom panel; see text for an explanation of these choices.}
    \label{fig:anievolz}
\end{figure*}

\begin{figure*}
	\includegraphics[height=3.3in]{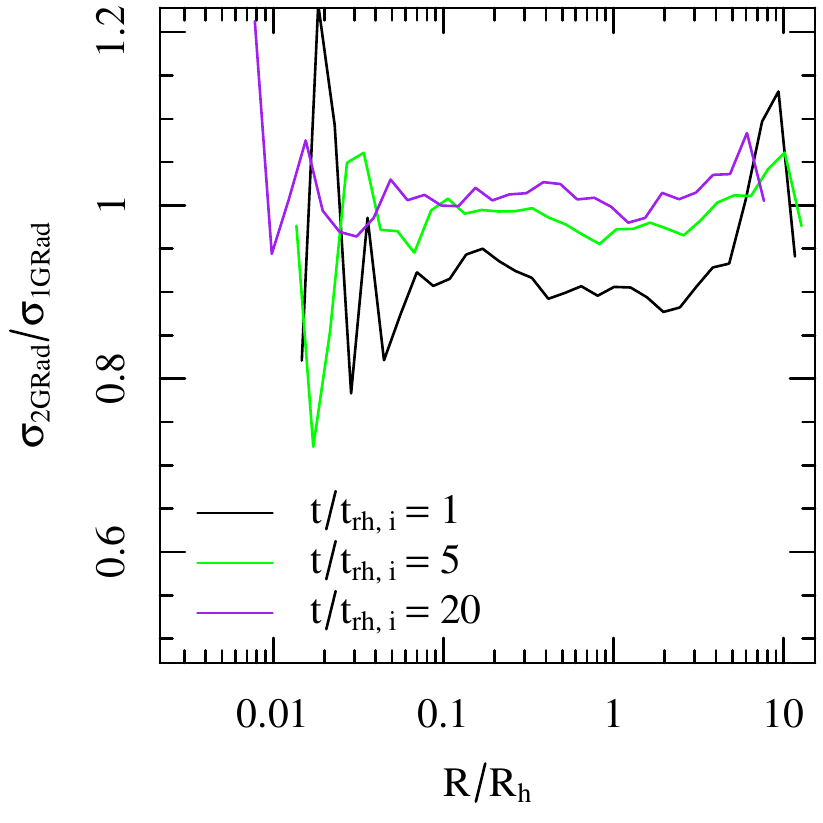}
    \includegraphics[height=3.3in]{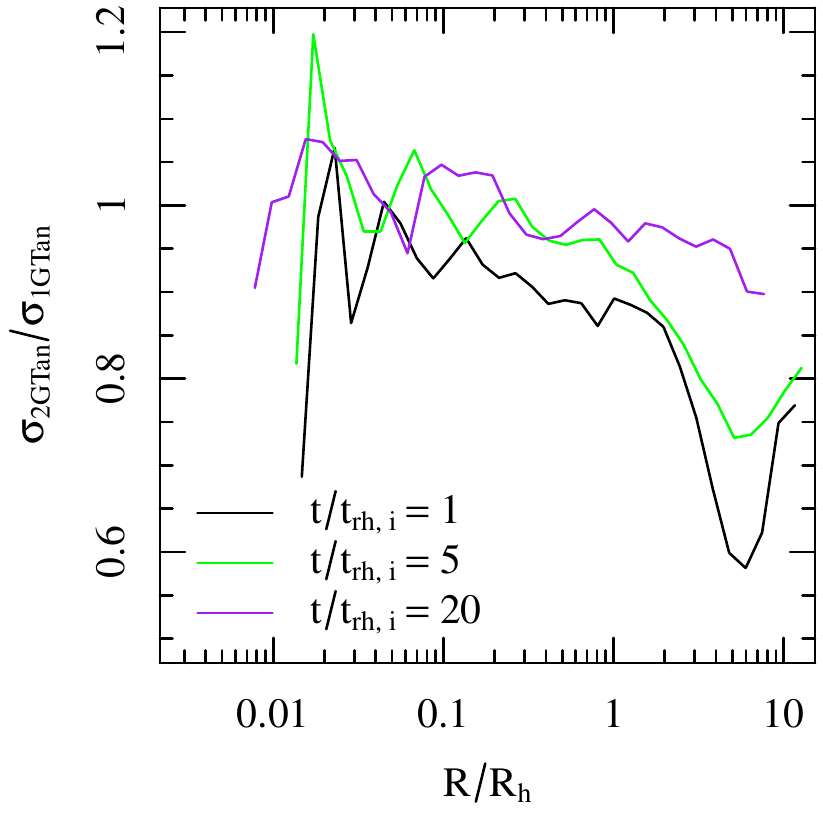}
    \caption{Ratio of the radial (left) and tangential (right) velocity dispersion profiles of each population in model Theta0, illustrated (at three different stages of evolution) as a function of radius normalised to the projected half-mass radius of the entire cluster model.}
    \label{fig:sigmarat}
\end{figure*}

We now focus our attention on the  evolution of the velocity anisotropy in our \Nbody models.  Fig.~\ref{fig:anievolz} shows the evolution of the velocity anisotropy profiles of multiple populations in models Theta0 and Theta45.
The initial conditions sampled from the \citet{varri2012} equilibria already possess some initial intrinsic velocity anisotropy that is strongest when observing the cluster along a line of sight parallel to the rotation axis \citep[see][for further details]{varri2012}. We first wish to focus on characterising the degree of evolutionary anisotropy (i.e., the one resulting from long-term, relaxation-driven effects). For this reason, we  choose a line of sight perpendicular to the rotation axis for the analysis of model Theta0.  On the other hand, for model Theta45 we adopt a generic direction as  the line of sight, to highlight the features associated with both evolutionary and initial intrinsic velocity anisotropy.
We define the anisotropy parameter as the ratio of the tangential velocity dispersion to the radial velocity dispersion, $\stsr$., both measured from projected velocities on the plane perpendicular to the line of sight.

Drawing parallels to the results of \citet{tiongco2016a}, we find that the evolution of the anisotropy profile of the 1G population is similar to that of a tidally filling system, while the evolution of the anisotropy of the more compact 2G subsystem is similar to that of a tidally underfilling case. Specifically, as the 2G population expands due to two-body relaxation, it populates the outer halo with 2G stars  on primarily radial orbits; this is clearly seen in Fig. \ref{fig:anievolz} as the 2G's evolution is characterised by a growing radial anisotropy in the cluster's outer regions. As shown in Fig.~\ref{fig:sigmarat}, where we compare the ratios of the tangential velocity dispersion profiles for each population, $\sigma_{\rm T,2G}/\sigma_{\rm T,1G}$ and also the ratios of the radial velocity dispersion profiles, $\sigma_{\rm R,2G}/\sigma_{\rm R,1G}$,  the larger radial anisotropy of the 2G population is due to the fact that this population has a smaller tangential velocity dispersion than the 1G. This behaviour is qualitatively consistent with that reported by \citet{richer2013,bellini2015,milone2018}, as resulting from state-of-the-art astrometric studies of several nearby Galactic globular clusters conducted with HST and Gaia DR2.
The less concentrated and more tidally filling 1G population never develops a strong radial anisotropy in the velocity space. As in the previous case, this behaviour is in agreement with the findings of \citet{tiongco2016a} for tidally filling systems.

As the system evolves and preferentially loses stars on more radial orbits \citep[see e.g.,][]{takahashi2000,baumgardt2003,tiongco2016a}, the overall radial anisotropy of the system decreases due to relaxation-driven star escape. Moreover, the magnitude of the strongest radial anisotropy decreases as  the Jacobi radius moves inwards (due to mass loss), stripping away the radially anisotropic layers and revealing the more isotropic layers closer to the centre of the cluster \citep[see][]{giersz1997}.  The loss of preferentially radial orbits also creates some tangential anisotropy in the outermost regions.

Similar to the case of the rotational velocity profiles, the two population's anisotropy profiles evolve to become increasingly similar as the two populations mix, after which the evolution proceeds similarly to that of a single population system. It is interesting to point out, however, that the two population may evolve to have a similar rotational profile while still be characterized by different anisotropy profiles (compare, for example, the anisotropy and rotation profiles for model Theta0: at $t=15\trhi$ in Figs~\ref{fig:vrotevol0} and \ref{fig:anievolz}).

\section{Discussion}
\label{discussion}

\begin{figure*}
	\includegraphics{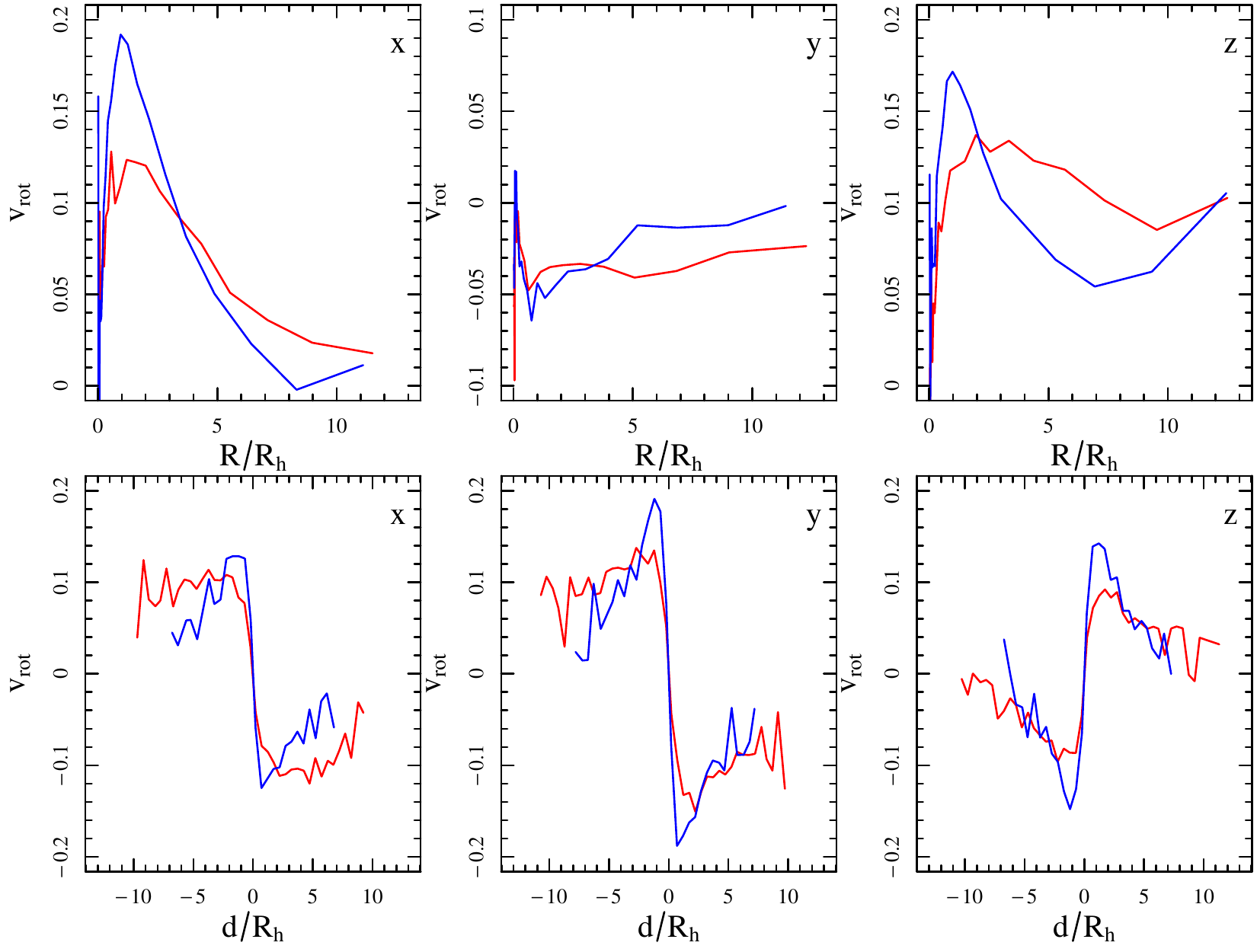}
    \caption{Rotation curves at time $t=1.5\trhi$ of each population in model Theta45, as calculated along three lines of sight, parallel to the $x$, $y$, and $z$-axes, respectively. The red coloured lines represent the 1G, and the blue lines represent the 2G.   Top panels: rotation curves derived from velocity components analogous to proper motion velocities, bottom panels: rotation curves derived from velocity components analogous to radial/line-of-sight velocities. All quantities are represented as a function of radius normalised to the projected half-mass radius of the entire cluster model.} 
    \label{fig:vrotlos}
\end{figure*}

\begin{figure*}
	\includegraphics{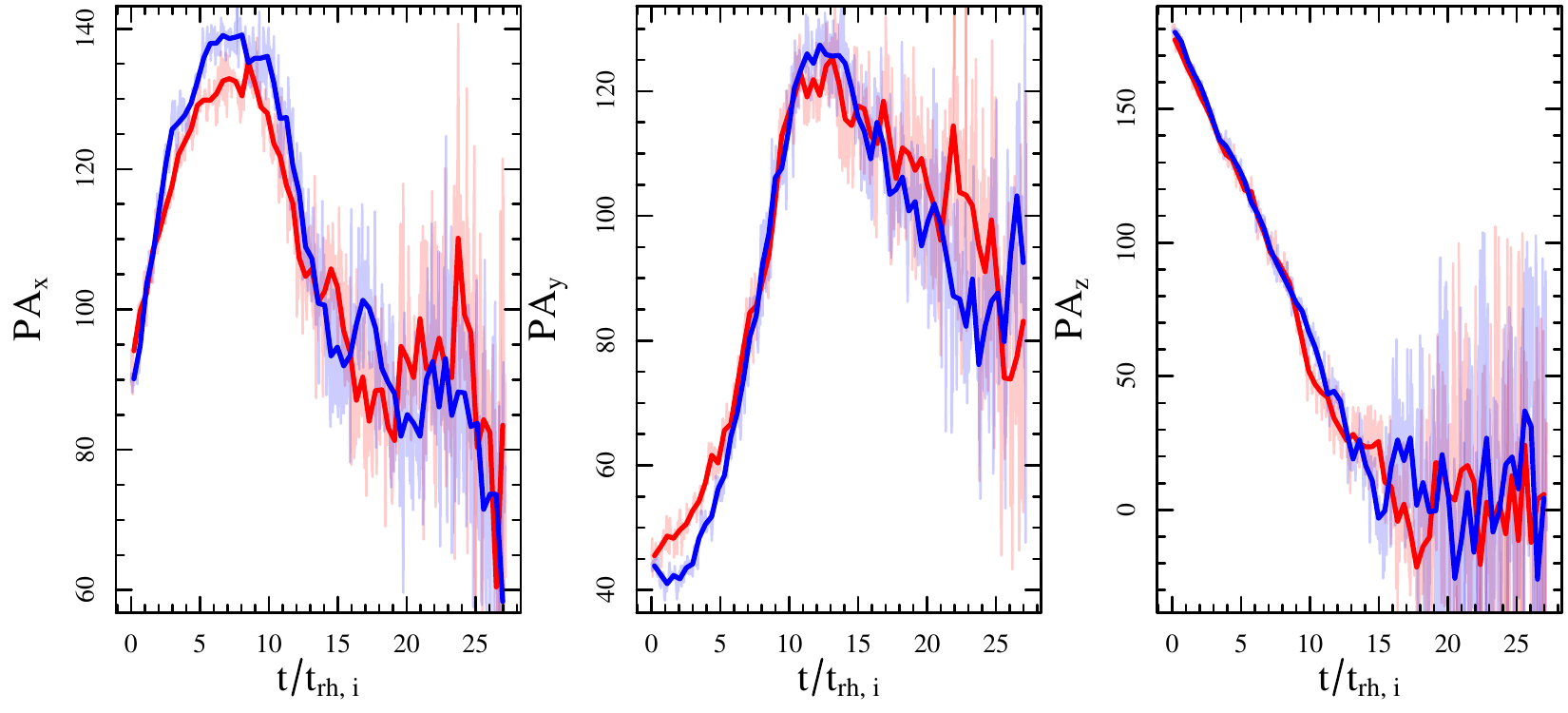}
    \caption{Position angles of the rotation axis for each population derived from velocities analogous to radial/line-of-sight velocities in model Theta45, as calculated along the three lines of sight adopted also in the analysis illustrated in Fig.~\ref{fig:vrotlos}. The red coloured lines represent the 1G, and the blue lines represent the 2G.  The lightly-shaded, thin solid lines are the PAs calculated at each snapshot, and the thick solid lines are the averaged values of PA across several  snapshots. All quantities are illustrated as a function of time, normalised to the initial value of the half-mass relaxation time of the entire cluster model. }
    \label{fig:paevol}
\end{figure*}

As discussed in the introduction, a growing number of observational studies have revealed that some Galactic globular clusters have retained some memory of the kinematical differences expected in scenarios predicting that 2G stars form more concentrated and more rapidly rotating than 1G stars.
The observational characterization of the kinematics of multiple populations, however, has been so far limited to a small set of cases and further studies will be essential to formulate further empirical constraints and a more complete picture of the dynamical dimension of the multiple population phenomenon.

On the theoretical side, the results of our direct \Nbody simulations have shown that the rich interplay between the dynamics of multiple-population formation, the dynamical processes affecting the cluster's long-term evolution and the cluster's interaction with the host galaxy's tidal field can lead to a number of complex features and non-trivial differences between the  kinematics of 1G and 2G populations. 
The awareness of these complexities has motivated the decision to focus our investigation exclusively on the interplay of only three fundamental physical ingredients: internal angular momentum, external tidal perturbations, and two-body relaxation. With such an illustrative goal in mind, the idealised \Nbody simulations presented in this study are appropriate to capture only the basic evolutionary processes at play. A detailed comparison with kinematic and photometric data of Galactic globular clusters would certainly require a careful exploration of a number of additional dimensions of the problem, both from the physical and numerical perspective, ranging from the effect of stellar evolution to the scaling with the number of particles. On the purely dynamical side, the presence of
a mass spectrum would introduce much greater complexity to the
definition of the problem (especially regarding the implications on
the process of transport of the angular momentum), which would
require a dedicated investigation. Finally, a number of degrees of
freedom associated with the design of the initial conditions (such
as the initial amount of angular momentum or the strength of the
external tidal field, as specified here by the initial filling factor and 
the orbital angular velocity) would also deserve a separate exploration, 
by means of a suite of dedicated \Nbody simulations.

Although our \Nbody simulations are still idealised and only aimed at exploring the fundamental dynamical processes taking place in multiple-population clusters, in this section we provide a few examples illustrating how the rich kinematical properties of the stellar systems we have explored in this study would (or would not) be revealed by data typically collected in current observational studies.
It is important to emphasise how the kinematic complexities emerging from the idealised star cluster models presented here and the possible challenges in their observational identification call for caution in the interpretation of observational studies based only on a partial characterisation of a cluster kinematics. Indeed, such caution further highlights the importance of building a complete picture based on the combination of line-of-sight velocities and proper motion measurements.

The lack of detection of kinematical differences can be due to the fact that in dynamically old clusters have effectively lost memory of all the fingerprints of the formation epoch and of those features usually produced during  dynamical evolution. However,  projection effects combined with the availability of only partial kinematical information may also limit our ability to identify existing kinematical differences or, possibly, reveal them but in a much weaker form.

Fig. \ref{fig:vrotlos} shows rotation curves of model Theta45 at $t=1.5\trhi$, as calculated along three different lines of sight. The top panels show the rotation curves calculated as in Section \ref{results}, equivalent to using proper motion data, and the bottom panels show the rotation curves calculated equivalent to using radial/line-of-sight velocities. The figure demonstrates that observing the cluster along different lines of sight will show different strengths in the rotation curve depending on how close the line of sight is to being parallel (for proper motions) or perpendicular (for radial velocities) to the actual rotation axis.  While the top panels show an obvious difference in the rotation curves of the two populations, the differences are less apparent using radial velocities.  Additionally, the rotation curves again show that one population doesn't necessarily rotate faster than the other in all regions of the cluster, i.e., the profiles show that 1G rotates faster than 2G in the outermost regions.  This representative example highlights the need for observations of the internal kinematics of multiple populations in globular clusters across a broad radial range, ideally throughout the entire cluster extension.

When calculating the rotation curve using radial velocities, the position angle (PA) of the rotation axis is chosen as the one that maximises the amplitude of $\langle\Delta v_{\rm r}\rangle$ \citep[see e.g.,][]{bellazzini2012}. However, in performing such an assessment, the size and radial distribution of the sample of stars under consideration can have a significant impact.  Often, all of the stars are taken into account, but previous observational and theoretical studies have shown that position angle of the rotation axis can vary as a function of the radial distance from the cluster centre \citep[see e.g.,][]{gebhardt2000,bianchini2013,boberg2017,tiongco2018}. Thus, in the case of multiple populations co-existing in a star cluster, if the populations are not yet spatially mixed, each population may also have a different position angle of its rotation axis; this effect has already been partly noted in M13 by \citet{cordero2017}.  In Fig.~\ref{fig:paevol}, we show the time evolution of the individual position angle of the rotation axis of each population, along three different lines of sight (this figure corresponds to Fig. \ref{fig:thetaomega}).  We find that the differences in PA are small enough so as not to make a significant difference in the rotation curves in each population. On the basis of this example, we conclude that, although possible misalignements of the PAs of the two populations appear to play only a secondary role in the characterisation of their corresponding kinematic observables, it is an aspect that, when possible, should always be empirically assessed, especially when interpreting observations of kinematical differences or lack thereof.

\section{Conclusions}
\label{conclusions}

We have presented the results of a set of direct $N$-body simulations aimed at understanding the fundamental aspects of the evolution of the internal kinematics of multiple stellar populations in tidally perturbed star cluster models.  We have followed the long-term dynamical evolution of multiple population clusters which, as suggested by hydrodynamical simulations of multiple-population formation, are composed of an extended, slower rotating first-generation (1G) population and a second-generation (2G) subsystem more spatially concentrated and more rapidly rotating than the 1G system in which it is embedded.

We focused our attention on the evolution of the degree of internal differential rotation and anisotropy in velocity space of the two populations, and we have explored their connection with the dynamical evolution of the cluster's structural properties and the effects of the host galaxy's tidal field.

We have studied the evolution of a tidally perturbed, rotating stellar system in which the orientation of the rotation axis is perpendicular to the plane of the cluster's  orbit around the host galaxy. We then considered the more general case of systems for which the initial rotation axis has a generic direction relative to orbital plane. In the latter case, as shown in \citet{tiongco2018}, a star cluster model can develop a number of complex kinematical properties resulting from the evolutionary interplay between relaxation-driven processes  and external influences by the tidal field.

In all cases, we find that during a cluster's long-term evolution, its overall internal rotation decreases due to the redistribution and loss of angular momentum, and that the rotation curves of the two populations become increasingly similar, eventually converging to a common rotational profile. 
The rotation curves of the two populations converge first in the inner regions of the cluster and subsequently in the outer regions. Before converging to a common behaviour, 2G stars are rotating more rapidly than the 1G component in the cluster's inner regions; however, in the cluster's outermost regions, where the 2G stars gradually move to as they diffuse from the inner regions, the 2G population is typically rotating more slowly than the 1G population.

Both populations eventually evolve towards a condition of internal rotation which is partially synchronized with the orbital motion, such that the internal angular velocity equals to half of the orbital angular velocity, as found in previous studies \citep{tiongco2016b,tiongco2017,claydon2017}. 

The results of the long-term dynamical evolution of multiple-population clusters starting with a rotation axis in a generic orientation relative the cluster's orbital angular velocity vector show a number of complex kinematical features.  As described in \citet{tiongco2018}, the orientation of cluster's global rotation axis varies with time due to precession and nutation oscillations induced by the cluster's interaction with the host galaxy's tidal field. The rotation axis direction may also vary with radius within the cluster: the outer regions are more strongly affected by the external tidal field and their kinematical properties evolve to acquire a rotation around a direction closer to the direction of  the cluster's orbital angular velocity vector. The inner regions, on the other hand, are less affected by the external tidal field and tend to maintain the direction of the initial intrinsic rotation. As a consequence of the initial differences in the spatial distributions of the two populations, the impact of the tidal effects on the internal kinematical properties is stronger for the less spatially concentrated 1G stars, and 1G and 2G stars develop a misalignment in their global angular velocity vectors and  different radial profiles of the direction of angular velocity vectors.

We also find significant differences in the velocity dispersion anisotropy of the two populations. The 2G system is initially compact, tidally underfilling, and develops a radial velocity anisotropy as 2G stars diffuse towards the cluster's outer regions.  The 1G system, on the other hand, is initially tidally filling and never develops a significant radial anisotropy. Differences in the anisotropy profiles of 1G and 2G can therefore be the kinematical manifestation of their different initial spatial properties.
Similar to the evolution of rotation profiles, the two populations' anisotropy profiles eventually become indistinguishable as the two populations mix.

Finally, we have discussed the possible observational limitations that can affect the ability to detect  the kinematical differences and complex features that we have found in our \Nbody simulations.  Projection effects may weaken or completely hide the signatures of the expected primordial or dynamically-induced kinematical differences. Moreover, the identification of many of the key fingerprints of the formation and dynamical evolution requires the ability to obtain  the variation of the 2G and 1G kinematical properties as functions of the radial distance from the cluster centre. These limitations should be  taken into account, and caution should be used in the interpretation of observational data.

The kinematical complexity emerging from our  star cluster models further emphasizes the need and importance of observational studies aimed at building a complete kinematical picture of the multiple population phenomenon, as based on a combination of radial and proper motion measurements, ideally covering the broadest possible range of distances from the cluster centre.

\section*{Acknowledgements}

We wish to thank the referee for a careful reading and review of the manuscript. 
This research was supported in part by Lilly Endowment, Inc., through its support for the Indiana University Pervasive Technology Institute, and in part by the Indiana METACyt Initiative. The Indiana METACyt Initiative at IU is also supported in part by Lilly Endowment, Inc.  MT acknowledges financial support from Indiana University's President's Diversity Dissertation Fellowship. ALV from a JSPS International Fellowship and Grant-in-Aid (KAKENHI-18F18787), a Marie Sklodowska-Curie Fellowship (MSCA-IF-EF-RI NESSY 658088), and the Institute for Astronomy at the University of Edinburgh.




\bibliographystyle{mnras}
\bibliography{main} 






\bsp	
\label{lastpage}
\end{document}